\documentclass[12pt]{article}
\usepackage{amssymb,amsmath,amsthm,latexsym}

\newtheorem{thm}{Theorem}[section]
\newtheorem{prop}[thm]{Proposition}
\newtheorem{lem}[thm]{Lemma}
\newtheorem{cor}[thm]{Corollary}
\newtheorem{exam}{Example}

\newtheorem{rem}[thm]{Remark}
\newcommand{\pf}{{\bf Proof. \ }}

\begin{document}

\title{Galois LCD Codes over Finite Fields}
\author{
 Xiusheng Liu\\
 School of Mathematics and Physics, \\
 Hubei Polytechnic University  \\
 Huangshi, Hubei 435003, China, \\
{Email: \tt lxs6682@163.com} \\
Yun Fan\\
 School of Mathematics and Statistics,\\
Central China Normal University,\\
Wuhan Hubei 430079, China,\\
{Email:\tt yfan@mail.ccnu.edu.cn}\\
Hualu Liu\\
School of Mathematics and Statistics,\\
Central China Normal University,\\
Wuhan Hubei 430079, China,\\
{Email: \tt hwlulu@aliyun.com} \\}
\maketitle


\begin{abstract}In this paper, we study the complementary dual  codes in more general setting (which are called Galois LCD codes) by a uniform method. A necessary and sufficient condition for linear codes  to be Galois LCD codes is determined, and  constacyclic  codes to be Galois LCD codes are characterized. Some illustrative examples which constacyclic  codes are Galois LCD MDS codes are provided as well. In particular, we study Hermitian LCD constacyclic codes. Finally, we present a construction of a class of Hermitian  LCD codes which are also MDS codes.
\end{abstract}


\bf Key Words\rm : Constacyclic codes, MDS  codes, cyclotomic cosets, Galois LCD  codes


\bf MSC2010\rm : 12E20, 94B60

\section{Introduction}

Linear complementary dual codes (which is abbreviated to LCD codes) are linear codes that meet their dual trivially. These codes were introduced by Massey in  \cite{Massey} and showed that asymptotically good LCD codes exist, and provided an optimum linear coding solution for the two-user binary adder channel. They are also used in counter measure to passive and active side channel analyses on embedded cryto-systems  \cite{Carlet}. Guenda, Jitman and Gulliver investigated an application of LCD codes in constructing good entanglement-assisted quantum error correcting codes \cite{Guend}.

Dinh established the algebraic structures in terms of generator polynomial of all repeated-root constacyclic codes of length $3p^s, 4p^s, 6p^s$ over finite field $\mathbb{F}_{p^m}$. Using these structures, constacyclic LCD codes of such lengths were also characterized (see \cite{Dinh10,Dinh13,Dinh132}). Yang and Massey in \cite{Yang} showed that a necessary and sufficient condition for a cyclic code of length $n$ over finite fields to be an LCD code is that the generator polynomial $g(x)$ is self-reciprocal and all the monic irreducible factors of $g(x)$ have the same multiplicity in $g(x)$ as in $x^n-1$. In \cite{Sendrier}, Sendrier indicated that linear codes with complementary-duals  meet the asymptotic Gilbert-Varshamov bound. Esmaeiliand Yari in \cite{Esmaeili} studied complementary-dual quasi-cyclic codes. Necessary and sufficient conditions for certain classes of quasi-cyclic codes to be LCD codes were obtained \cite{Esmaeili}. Dougherty, Kim, Ozkaya, Sok and Sol$\acute{e}$ developed a linear programming bound on the largest size of an LCD code of given length and minimum distance \cite{Dougherty}. In recently, Ding, C. Li and S. Li in \cite{Ding} constructed  LCD BCH codes.  In addition, Boonniyoma and Jitman gave a study on linear codes with Hermitian complementary dual \cite{Boonniyoma}, and we also in \cite{Liu} studied LCD codes over finite chain rings.

Constacyclic  codes  over finite fields  are important classes of linear codes in theoretical and practical viewpoint. In \cite{Fan}, Fan and Zhang studied  Galois self-dual constacyclic codes over finite fields. Motivated by this work, we will investigate Galois  complementary dual codes (which is abbreviated to Galois LCD codes) over finite fields. Some of them have better parameters.

In this work, we study the complementary dual  constacyclic codes in more general setting  by a uniform method. The necessary background materials of Galois dual and the definition of Galois LCD codes are given in Section 2. Moreover, we obtain a criteria of Galois LCD codes. In Section 3, we characterize the generator polynomials of Galois LCD constacyclic  codes. Next, we obtain a sufficient and necessary condition for a code $C$ to be an Galois LCD constacyclic  code over finite fields and give examples that $C$ is an Galois LCD MDS constacyclic over finite fields. Finally, in Section 4, we  address the Hermitian LCD  constacyclic codes over finite fields and get a family of Hermitian LCD MDS  constacyclic codes.

\section{ Galois  LCD codes over $\mathbb{F}_{q}$}
Throughout this paper, we denote by  $\mathbb{F}_{q}$ the finite field with cardinality $\mid\mathbb{F}_{q}\mid=q=p^e$, where $p$  is a prime and $e$ is a positive integer. Let $\lambda\in \mathbb{F}_{q}^{*}$, where $\mathbb{F}_{q}^{*}$ denotes the multiplicative group of units of $\mathbb{F}_{q}$, and let $n$ be a positive integer coprime to $q$. Any ideal $C$ of the quotient ring $R_{n,\lambda}=\mathbb{F}_{q}[X]/\langle X^n-\lambda\rangle$ is said to be a $\lambda$-constacyclic code of length $n$ over $\mathbb{F}_{q}$.
Let $\mathbb{F}_{q}^n=\{{\bf x}=(x_1,\cdots,x_n)\,|\,x_j\in \mathbb{F}_{q}\}$ be $n$ dimensional vector  space over
$\mathbb{F}_{q}$. A subspace $C$ of $\mathbb{F}_{q}^n$ is called a linear code
of length $n$ over $\mathbb{F}_{q}$. We assume that all codes are
linear. If  a linear code $C$ over $\mathbb{F}_{q}$ with parameters $[n,k,d]$ attains the  Singleton bound $d= n-k+1$, then it is called a maximum-distance-separable (MDS) code.

Let ${\bf x},{\bf y}\in\mathbb{F}_{q}^n$. In \cite{Fan}, Fan and Zhang introduce a kind of inner products,  called Galois inner product, as follows: for each integer $k$ with $0\leq k< e $ , define:
$$
[{\bf x},{\bf y}]_{k}=x_1y_1^{p^{k}}+\cdots+x_ny_n^{p^{k}}.
$$
It is just the usual Euclidean inner product if $k=0$. And, it is  the Hermitian inner product if $e$ is even and $k=\frac{e}{2}$.
we call
$$
C^{\bot_{k}}=\{{\bf x}\in\mathbb{F}_{q} ^n\,|[{\bf c},{\bf x}]_{k}=0,\forall{\bf c}\in C\}
$$
as the Galois dual code of $C$.  It is easy to see that $C^{\bot_{0}}$ (simply, $C^{\perp}$) is just the Eucilidean dual code of $C$ and $C^{\bot_{\frac{e}{2}}}$ (simply, $C^{\perp_{H}}$) is just the Hermitian dual code of $C$.

Notice that $C^{\bot_{k}}$ is linear if
$C$ is linear or not.

From the fact that Galois inner product is nondegenerate, it
follows immediately that  $\mathrm{dim}_{\mathbb{F}_{q}}C+\mathrm{dim}_{\mathbb{F}_{q}}C^{\perp_{k}}=n$.

A  linear code $C$ is called Galois LCD  if $C^{\perp_{k}}\cap C=\{\mathbf{0}\} $, and an Galois LCD code $C$ is called Galois LCD  MDS if $C$ attains the Singleton bound.

Given a vector $\mathbf{a}=(a_1,a_2,\ldots,a_n)\in\mathbb{F}_{q} ^n$,  we  define the $p^{e-k}$th power of $\mathbf{a}$ as
$$\mathbf{a}^{p^{e-k}}=(a_1^{p^{e-k}},a_2^{p^{e-k}},\ldots,a_n^{p^{e-k}}).$$
For a linear code $C$ of length $n$ over $\mathbb{F} _{q}^{n}$, we define $C^{p^{e-k}}$ to be the
set $\{\mathbf{a}^{p^{e-k}}\mid~\mathrm{for}~\mathrm{all}~\mathbf{a}\in C\}$.  Then,  it is easy to see that for a linear code $C$ of length $n$ over $\mathbb{F} _{q}^{n}$, the Galois dual $C^{\perp_{k}}$ is equal to the Euclidean $(C^{p^{e-k}})^{\perp}$ dual of $C^{p^{e-k}}$.

Let $A =(a_{ij} )$ be an $s\times s$ matrix with entries in $\mathbb{F}_{q}$ , we define $A^{(p^{e-k})} = (a_{ij}^{p^{e-k}})$. The following lemma is clear.
\begin{lem} If $C$ is an $[n,l,d]$ linear code  over $\mathbb{F}_{q}$ with a generating matrix $G$, then $C^{p^{e-k}}$ is also an $[n,l,d]$ linear code  over $\mathbb{F}_{q}$ with a generating matrix $G^{(p^{e-k})}$. Moreover, $C$ is Galois LCD if and only if $C \cap (C^{p^{e-k}})^{\perp}=\{\mathbf{0}\}$.
\end{lem}

The following theorem gives a criteria of Galois LCD codes and is analogous to the result of Eucilidean LCD codes in \cite{Massey}.
\begin{thm} Let C be an $[n,l,d]$ linear code over $\mathbb{F} _{q}$ with generator matrix
$G$. Then $C$ is  Galois LCD if and only if $G(G^{(p^{e-k})})^{T}$ is nonsingular.
\end{thm}
\pf Suppose that $G(G^{(p^{e-k})})^{T}$ is singular. Then there exists a nonzero $\mathbf{a}\in \mathbb{F} _{q}^{n}$ such that $\mathbf{a}G(G^{(p^{e-k})})^{T}=\mathbf{0}$. Taking $\mathbf{c}=\mathbf{a}G$, it is clear that $ \mathbf{c}\in C\backslash\{\mathbf{0}\}$ and it satisfies $\mathbf{c}(G^{(p^{e-k})})^{T}=\mathbf{0}$. It follows that $\mathbf{c}\in (C^{p^{e-k}})^{\perp}$, which is a contraction.

Conversely, assume that $G(G^{(p^{e-k})})^{T}$ is nonsingular. Let $\mathbf{u}\in\mathbb{F} _{q}^{n}$. If
$\mathbf{u}\in C$, then there
exists $\mathbf{v}\in\mathbb{F} _{q}^{l}$ such that $\mathbf{u}=\mathbf{v}G$. It follows that
$$\mathbf{u}(G^{(p^{e-k})})^{T}(G(G^{(p^{e-k})})^{T})^{-1}G=\mathbf{v}G(G^{(p^{e-k})})^{T}(G(G^{(p^{e-k})})^{T})^{-1}G=\mathbf{v}G=\mathbf{u}.$$
If $\mathbf{u}\in C^{\perp_{k}}$, then $\mathbf{u}(G^{(p^{e-k})})^{T}=\mathbf{0}$, and hence
$$\mathbf{u}(G^{(p^{e-k})})^{T}(G(G^{(p^{e-k})})^{T})^{-1}G=\mathbf{0}(G(G^{(p^{e-k})})^{T})^{-1}G=\mathbf{0}.$$

For any $\mathbf{a}\in C\cap C^{\perp_{k}}$, by $\mathbf{a}\in C$, we have $\mathbf{a}=\mathbf{a}(G^{(p^{e-k})})^{T}(G(G^{(p^{e-k})})^{T})^{-1}G$, and by $\mathbf{a}\in C^{\perp_{k}}$ again, we have also $\mathbf{a}=\mathbf{a}(G^{(p^{e-k})})^{T}(G(G^{(p^{e-k})})^{T})^{-1}G=\mathbf{0}$. Therefore, $C\cap C^{\perp_{k}}=\{\mathbf{0}\}$, i.e., $C$ is Galois LCD.
\qed

It is well known that, for a given $[n,l, d]$ code over $\mathbb{F} _{q}$, there exists an equivalent code
with the same parameters such that its generator matrix is of the form $G=[I_l~A]$
for some $l\times(n-l)$ matrix $A$ over $\mathbb{F} _{q}$, where $I_l$ is a $l\times l$ identity matrix. The generator matrix of a linear code of this
form plays an important role in constructing Galois LCD codes.

The following fact is well known.
\begin{lem} Let $p$ be a prime. If $p\equiv 1~ \mathrm{mod ~4}$, then $-1$ is a quadratic modulo $p$.
\end{lem}

\begin{thm}  Let $\widetilde{C}$ be an $[n,l,\widetilde{d}]$ linear code over $\mathbb{F} _{q}$
with generator matrix $\widetilde{G}=[I_l~A]$.

$\mathrm{(1)} $ If $\mathrm{char}\mathbb{F} _{q} = 2$, then exists a Galois LCD code $C$ over $\mathbb{F} _{q}$ with parameters $[2n-l,l,d]$ and $d\geq \widetilde{d}$.

$\mathrm{(2)} $ If $\mathrm{char}\mathbb{F} _{q} \equiv 1 ~\mathrm{mod} ~4$, then there exists $\eta\in \mathbb{F} _{q}$  such that $\eta^2=-1$ and a
linear code $C$ generated by $G= [ I_l~A~\eta A]$ is a Galois LCD code over $\mathbb{F} _{q}$ with parameters $[2n-l,l,d]$ and $d\geq \widetilde{d}$.
\end{thm}
\pf $\mathrm{(1)} $ When $\mathrm{char}\mathbb{F} _{q} = 2$. Let $C$ be a linear code generated by $G=[I_l~A~A]$ over $\mathbb{F} _{q}$. Then
 $$G(G^{(p^{e-k})})^{T}=I_l+A(A^{(p^{e-k})})^{T}+A(A^{(p^{e-k})})^{T}=I_l.$$
Therefore, $G(G^{(p^{e-k})})^{T}$ is nonsingular,  which implies code $C$ is Galois LCD.

Next, we show that $d(C)\geq \widetilde{d} $. Let  $\mathbf{u} \in C \backslash \{\mathbf{0}\}$. Then there exists $ \mathbf{v} \in \mathbb{F} _{q}^{l} \backslash \{\mathbf{0}\}$ such that $\mathbf{u}=\mathbf{v}G=[\mathbf{v}I_l~\mathbf{v}A ~\mathbf{v}A]$. Hence,
$$ W_H(\mathbf{u})=W_H([\mathbf{v}I_l~\mathbf{v}A ~\mathbf{v}A])\geq W_H([\mathbf{v}I_l~\mathbf{v}A])=W_H(\mathbf{v}(\widetilde{G}))\geq \widetilde{d},$$
which implies $d\geq \widetilde{d}$.

$\mathrm{(2)} $ When $\mathrm{char}\mathbb{F} _{q} \equiv1 ~\mathrm{mod} ~4$. For $0\leq k<e$, we can assume $p^{e-k}=4t+1$ where $t$ is an integer. Then $\eta^{1+p^{e-k}}=\eta^{2(2t+1)}=-1$. Therefore,
$$G(G^{(p^{e-k})})^{T}=[I_l~A~\eta A][I_l~A^{(p^{e-k})}~\eta^{p^{e-k}} A^{(p^{e-k})}]^{T}=I_l+A(A^{(p^{e-k})})^{T}+\eta^{p^{e-k}+1}A(A^{(p^{e-k})})^{T}=I_l.$$
This means that $C$ is a Galois LCD code over $\mathbb{F} _{q}$.

Similar to $\mathrm{(1)} $, we can prove that $C$ is an $[2n-l, l,d]$ code with $d\geq \widetilde{d}$.

\begin{exam} Let $C$ be a linear code of length $4$ over $\mathbb{F}_{8}=\{0,1=\alpha^7=\alpha^0,\alpha,\alpha^2,\alpha^3=1+\alpha,\alpha^4=\alpha+\alpha^2,\alpha^5=1+\alpha+\alpha^2,\alpha^6=1+\alpha^2\}$
with generator matrix $G=\begin{pmatrix}1&0&\alpha&\alpha\\0&1&1&\alpha\end{pmatrix}$. Take $k=1$. Then $p^{e-k}=2^{3-1}=4$. Since $\mathrm{det}[G(G^{(4)})^{T}]=\alpha\neq0$, we have $G(G^{(4)})^{T}$ is  nonsingular. Hence, $C$ is a Galois LCD MDS code over $\mathbb{F}_{8}$ with parameters $[4,2,3]$.
\end{exam}

\section {Galois LCD constacyclic codes over $\mathbb{F}_{q}$}
In this section, we investigate Galois LCD $\lambda$-constacyclic codes over $\mathbb{F}_{q}$. The following proposition in \cite{Dinh10}  is very usual.

\begin{prop} Let $\alpha,\beta$ be distinct nonzero elements of the field  $\mathbb{F}_{q}$.  Then  a linear code $C$ of length $n$ over $\mathbb{F}_{q}$ is both $\alpha$-and $\beta$-constacyclic  if and only if $C=\{\mathbf{0}\}$ or  $C=\{\mathbb{F}_{q}^{n}\}$.
\end{prop}

The following lemma can be found in \cite{Fan}.
\begin{lem} If $C$ is a $\lambda$-constacyclic code of length $n$ over $\mathbb{F}_{q}$, then $C^{\perp_{k}}$ is a $\lambda^{-p^{e-k}}$-constacyclic code of length $n$ over $\mathbb{F}_{q}$ .
\end{lem}

\begin{cor} If $\lambda^{1+p^{e-k}}\neq1$, then any $\lambda$-constacyclic $C$ of length $n$ over $\mathbb{F}_{q}$ is a Galois LCD code.
\end{cor}
\pf Indeed, by Lemma 3.2, if $C$ is a $\lambda$-constacyclic code then $C^{\perp_{k}}$ is a $\lambda^{-p^{e-k}}$-constacyclic code. Thus, $C\cap C^{\perp_{k}}$ is both $\lambda$-and $\lambda^{-p^{e-k}}$-constacyclic. When $\lambda^{1+p^{e-k}}\neq1$, as  $C\cap C^{\perp_{k}}$ can not be $\mathbb{F}_{q}^{n}$, by Proposition 3.1, $C\cap C^{\perp_{k}}=\{\mathbf{0}\}$, i.e., $C$ is a Galois LCD code.
\qed

By Corollary 3.3, when $\lambda^{1+p^{e-k}}\neq1$, any $\lambda$-constacyclic code $C$ is  a Galois  LCD code. Thus, in order to obtain all  Galois LCD $\lambda$-constacyclic  codes, we only need to look at the  classes of  $\lambda$-constacyclic codes where $\lambda^{1+p^{e-k}}=1$.

We first give the definition of reciprocal polynomial in $\mathbb{F}_q[x]$.  Then we study the generator polynomials of  Galois  LCD   $\lambda$-constacyclic  codes.

For a polynomial $f(x)=\sum_{i=0}^{l}a_{i}x^{i}$ of degree $l$ ($a_{0}\neq0)$ over $\mathbb{F}_{q}$, let $\widetilde{f}(x)$ denote the monic reciprocal polynomial of $f(x)$ given by
\\$$\widetilde{f}(x)=a_{0}^{-1}x^{l}f(\frac{1}{x})=a_{0}^{-1}\sum_{i=0}^{l}a_{i}x^{l-i}.$$

It is well-known that a nonzero $[n,l]$ $\lambda$-constacyclic code $C$ has a unique generator polynomial $g(x)$ of degree $n-l$, where $g(x)|x^n-\lambda$. The roots of the code $C$ are the roots of $g(x)$. So, if $\xi_1,\ldots,\xi_{n-l}$ are the roots of $g(x)$  in some  extension field of $\mathbb{F}_{q}$, then $\mathbf{c}=(c_0,c_1,\ldots,c_n)\in C$ if and only if $c(\xi_1)=\cdots=c(\xi_{n-l})=0$, where $c(x)=c_0+c_1x+\cdots c_{n-1}x^{n-1}$. Let $h(x)=\frac{x^n-\lambda}{g(x)}=\sum_{i=0}^{l}h_ix^i$. Then we have the following lemma.
\begin{lem} With notations as above. Let $C$ be an $\lambda$-constacyclic code of length $n$ over $\mathbb{F}_{q}$. Then

$\mathrm{(1)}$ $C^{\perp}$ is an $\lambda^{-1}$-constacyclic code generated by $\widetilde{h}(x)=\sum_{i=0}^{l}h_0^{-1}h_ix^{l-i}$;

$\mathrm{(2)}$ $C^{\perp_{k}}$ is an $\lambda^{-p^{e-k}}$-constacyclic code generated by $\widetilde{h}^{p^{e-k}}(x)=\sum_{i=0}^{l}h_0^{-p^{e-k}}h_i^{p^{e-k}}x^{l-i}$.
\end{lem}
\pf $\mathrm{(1)}$ The proof can be found \cite{Dinh20}.

$\mathrm{(2)}$ Set $\widetilde{C}=\langle \widetilde{h}^{p^{e-k}}(x)\rangle$. Suppose that $\eta_1,\ldots,\eta_l$ be the zeros of  $\widetilde{h}(x)$. Then
$\eta_1^{p^{e-k}},\ldots,\eta_l^{p^{e-k}}$  are the zeros of $\widetilde{h}^{p^{e-k}}(x)$. This means that if $(c_0,c_1,\ldots,c_n)\in C^{\perp}$,  then  $(c_0^{p^{e-k}},\ldots,c_{n-1}^{p^{e-k}})\in \widetilde{C}$.

The following we first prove that
$$(C^{\perp})^{p^{e-k}}=(C^{p^{e-k}})^{\perp}.~~~~~~~~~~~~~~~~~~~~~~~~~~~~~~$$
Suppose $G$ is a generator matrix of $C$. It is easy to prove that $G^{p^{e-k}}$ is a generator matrix of $C^{p^{e-k}}$.

Similarly, if $H$ is a parity-check for $C$,  then $H^{p^{e-k}}$ is a generator matrix of $(C^{\perp})^{p^{e-k}}$.

Suppose that $G=\begin{pmatrix}g_1\\g_{2}\\ \vdots \\ g_{k}\end{pmatrix}$ and $H=\begin{pmatrix}h_1\\h_{2}\\ \vdots \\ h_{n-k}\end{pmatrix}$. Then $G^{p^{e-k}}=\begin{pmatrix}g_1^{p^{e-k}}\\g_{2}^{p^{e-k}}\\ \vdots \\ g_{k}^{p^{e-k}}\end{pmatrix}$ and
$H^{p^{e-k}}=\begin{pmatrix}h_1^{p^{e-k}}\\h_{2}^{p^{e-k}}\\ \vdots \\ h_{n-k}^{p^{e-k}}\end{pmatrix}$.

For $y\in(C^{\perp})^{p^{e-k}}$, we can assume that
$$y=t_{1}h_{1}^{p^{e-k}}+\cdots+t_{n-k}h_{n-k}^{p^{e-k}}.$$
Then for any $g_{j}^{p^{e-k}}\in G^{p^{e-k}}$, one obtain that
$$[y,g_{j}^{p^{e-k}}]=\sum_{i=1}^{n-k}t_{i}[h_{i}^{p^{e-k}},g_{j}^{p^{e-k}}]=\sum_{i=1}^{n-k}t_{i}[h_{i},g_{j}]^{p^{e-k}}=0.$$
Therefore, $y\in(C^{p^{e-k}})^{\perp}$, which implies that $(C^{\perp})^{p^{e-k}}\subset(C^{p^{e-k}})^{\perp}$.

On the other hand, we verity that if $v_{1},v_{2}, \ldots, v_{k}$ are linear independent vectors in $\mathbb{F}_{q}$, then  $v_{1}^{p^{e-k}},v_{2}^{p^{e-k}}, \ldots, v_{k}^{p^{e-k}}$ are also linear independent vectors in $\mathbb{F}_{q}$. In fact, assume that
$$a_{1}v_{1}^{p^{e-k}}+a_2v_{2}^{p^{e-k}}+\cdots+a_kv_{k}^{p^{e-k}}=0,$$
where $a_{1},a_2,\cdots,a_k\in\mathbb{F}_{q}$, then
$$a_{1}^{p^k}v_{1}+a_2^{p^k}v_{2}+\cdots+a_k^{p^k}v_{k}=0.$$
Since $v_{1},v_{2}, \ldots, v_{k}$ are linear independent vectors in $\mathbb{F}_{q}$, $a_{1}^{p^k}=a_2^{p^k}=\cdots=a_k^{p^k}=0$. Hence  $a_{1}=a_2=\cdots=a_k=0$. This meant that $v_{1}^{p^{e-k}},v_{2}^{p^{e-k}}, \ldots, v_{k}^{p^{e-k}}$ are linear independent vectors in $\mathbb{F}_{q}$.

According to the fact above proof, we have
$$~~~~~~~~~~\mathrm{dim}(C^{\perp})^{p^{e-k}}=n-k=\mathrm{dim}(C^{p^{e-k}})^{\perp}.~~~~~~~~~~~~~~~~$$
Summarizing, we have $(C^{\perp})^{p^{e-k}}=(C^{p^{e-k}})^{\perp}$. Thus
$$C^{\perp_{k}}=(C^{p^{e-k}})^{\perp}=\{(c_0^{p^{e-k}},\ldots,c_{n-1}^{p^{e-k}})|(c_0,c_1,\ldots,c_n)\in C^{\perp}\}.$$
It follows that $C^{\perp_{k}}\subset \widetilde{C}$. Since $\mathrm{dim}_{\mathbb{F}_{q}}C^{\perp_{k}}=\mathrm{dim}_{\mathbb{F}_{q}}\widetilde{C}$, we get $C^{\perp_{k}}= \widetilde{C}$.
\qed

The following gives  a criteria of Galois LCD $\lambda$-constacyclic  codes. We first take the following notations:

$\bullet$ $\mathrm{ord}_{\mathbb{F}_{q}^{*}}(\lambda)=r$, where $\mathrm{ord}_{\mathbb{F}_{q}^{*}}(\lambda)$  denotes the order of $\lambda$ in multiplicative group;

$\bullet$ $\mathbb{Z}_{rn}$ denotes the residue ring of the integer ring $\mathbb{Z}$ modulo $rn$;

$\bullet$ $\mathbb{Z}_{rn}^{*}$ denotes the  multiplicative group consisting of units of  $\mathbb{Z}_{rn}$;

$\bullet$ $1+\mathbb{Z}_{rn}=\{1+rt|t=0,1,\ldots,n-1\}\subset \mathbb{Z}_{rn}$;

$\bullet$ $\mu_{s}$, where $\mathrm{gcd}(s,rn)=1$,  denotes the permutation of the set  $\mathbb{Z}_{rn}$ given by $\mu_{s}(x)=sx$ for all $x\in \mathbb{Z}_{rn}$.

Let $m$ be the  multiplicative order of $q$ modulo $rn$, i.e., $rn\mid(q^m-1)$ but $rn\nmid(q^{m-1}-1)$. Then, in $\mathbb{F}_{q^m}$,  there exists a primitive $rn$th root $\theta$ of unity such that $\theta^{n}=\lambda$. It is easy to check that $\theta^{i}$ for all $i \in (1+\mathbb{Z}_{rn})$ are all roots of $x^n-\lambda$. In $\mathbb{F}_{q^m}[x]$,  we have the following decomposition:
$$x^n-\lambda=\prod_{i\in(1+\mathbb{Z}_{rn})}(x-\theta^{i}).$$

Since $\mathrm{gcd}(q,n)=1$ and $r\mid( q-1)$, it follows that $q\in \mathbb{Z}_{rn}^{*}\cap(1+\mathbb{Z}_{rn})$ and $1+\mathbb{Z}_{rn}$ is $\mu_q$-invariant. Let $(1+\mathbb{Z}_{rn})/\mu_q$ denote the set of $ \mu_q$-orbits on $1+\mathbb{Z}_{rn}$, i.e., the set of $q$-cyclotomic cosets on $1+\mathbb{Z}_{rn}$. For any $q$-cyclotomic coset $Q$ on $1+\mathbb{Z}_{rn}$, the polynomial $ M_Q(x)=\prod_{i\in Q}(x-\theta^i)$ is irreducible in $\mathbb{F}_{q}[x]$. We  further get a monic irreducible decomposition as follows:
$$x^n-\lambda=\prod_{Q\in(1+\mathbb{Z}_{rn})/\mu_{q}}M_Q(x).$$

The defining set of the  $\lambda$-constacyclic code $C$ is defined as
$$P=\{1+ir\in(1+\mathbb{Z}_{rn})|\theta^{1+ir} ~\mathrm{is\\~a~root~of}~C\}.$$
It is clearly to see that $P$ is a union of some  $q$-cyclotomic cosets modulo $rn$ and  $\mathrm{dim}C=n-\mid P\mid$.

Similar to cyclic codes, there exists the following BCH bound for constacyclic codes (see\cite{Bocong}).

\begin{thm} (The BCH bound for constacyclic codes) Suppose that $gcd(q,n)=1$. Let $C=\langle g(x)\rangle$ be an  $\lambda$-constacyclic code of length $n$ over $\mathbb{F}_{q}$ with the roots $ \{\theta^{1+ri}|i_1\leq i\leq i_1+d-1\}$. Then the minimum distance of $C$ is at least $d$.
\end{thm}

\begin{lem} If $\lambda^{1+p^{e-k}}=1$, then $-p^{e-k}(1+r\mathbb{Z}_{rn})=1+r\mathbb{Z}_{rn}~(\mathrm{mod}~rn)$.
\end{lem}
\pf Since $\lambda^{1+p^{e-k}}=1$ and $\mathrm{ord}_{\mathbb{F}_{q}^{*}}(\lambda)=r$, we have $r\mid(1+p^{e-k}) $. Suppose that $1+p^{e-k}=rt$. Then for $1+ir\in (1+\mathbb{Z}_{rn})$ we have
$$-p^{e-k}(1+ir)=-p^{e-k}ir-(p^{e-k}+1)+1=(-p^{e-k}i-t)r+1~(\mathrm{mod}~rn)\in(1+\mathbb{Z}_{rn})$$
Thus, we obtain $-p^{e-k}(1+r\mathbb{Z}_{rn})=1+r\mathbb{Z}_{rn}~(\mathrm{mod}~rn)$.
\qed

\begin{thm} Let $C_P$ be an   $\lambda$-constacyclic code of length $n$ over $\mathbb{F}_q$ with the defining set $P$, where $\lambda^{1+p^{e-k}}=1$. Let $\overline{P}=(1+\mathbb{Z}_{rn})\setminus P$. Then

$\mathrm{(1)}$ $-p^{e-k}\overline{P}$  is the defining set of a Galois dual code of $C_P$, i.e., $-p^{e-k}\overline{P}$  is the defining set of the $\lambda^{-p^{e-k}}$-constacyclic code  $C_P^{\perp_{k}}$.

$\mathrm{(2)}$ $C_P$ is a Galois LCD $\lambda$-constacyclic  code if and only if $-p^{k}P=P$.
\end{thm}
\pf $\mathrm{(1)}$ According to Lemma 3.6, $-p^{e-k}\overline{P}\subset (1+\mathbb{Z}_{rn})$.  It is easy to see that $-p^{e-k}\overline{P}$ is a union of some $q$-cyclotomic cosets containing in $(1+\mathbb{Z}_{rn})$ and $\mid P\mid+\mid -p^{e-k}\overline{P}\mid=n$.

It is clear that
$$x^n-\lambda=\prod_{i\in(1+\mathbb{Z}_{rn})}(x-\theta^{i})=\prod_{i\in P}(x-\theta^{i})\prod_{i \in (-\overline{P})}(x-\theta^{i}).$$

By Lemma 3.4(2), the generator polynomial of $\lambda^{-p^{e-k}}$-constacyclic code  $C_P^{\perp_{k}}$ is
$$\widetilde{h}^{p^{e-k}}(x)=h_0^{-p^{e-k}}x^l\prod_{i\in(-\overline{P})}(\frac{1}{x}-\theta^{-i})^{p^{e-k}}=\prod_{i\in(-\overline{P})}(x-\theta^{ip^{e-k}})=\prod_{j\in(-p^{e-k}\overline{P})}(x-\theta^{j}).$$
Thus,  $-p^{e-k}\overline{P}$  is the defining set of the $\lambda^{-p^{e-k}}$-constacyclic code  $C_P^{\perp_{k}}$.

 $\mathrm{(2)}$ Let $f_{\overline{P}}(x)=\prod_{i\in\overline{P}}(x-\theta^{i})$. Then, according to the definition of $\overline{P}$, $f_{\overline{P}}(x)$ is check polynomial of $C_P$.

 Similarly, let $f_{-p^{e-k}P}(x)=\prod_{i\in (-p^{e-k}P)}(x-\theta^{i})$. Then, by $\mathrm{(1)}$, $f_{-p^{e-k}P}(x)$ is check polynomial of $C_P^{\perp_{k}}$.
Therefore, $C_P\cap C_P^{\perp_{k}}=\{\mathbf{0}\}$ if and only if $\overline{P}\cap (-p^{e-k}P)=\phi$, i.e., $C_P\cap C_P^{\perp_{k}}=\{\mathbf{0}\}$ if and only if  $-p^{e-k}P=P$. This means that $C_P$ is a Galois LCD $\lambda$-constacyclic  code if and only if $-p^{k}P=P$.
\qed

\begin{exam} Let $p=11,e=3,i.e., q=11^3=1331$.Take $k=1$, $n=5$ and $\lambda=-1$. Then $r=2,rn=10$.  Consider
 $$1+2\mathbb{Z}_{10}=\{1,3,5,7,9\}.$$
 The $q$-cyclotomic cosets modulo $10$ containing in $1+2\mathbb{Z}$ are
 $$Q_1=\{1\},Q_3=\{3\},Q_5=\{5\},Q_7=\{7\},Q_{9}=\{9\}.$$
 It is easy to check that $-11Q_1=Q_9,-11Q_9=Q_1,-11Q_3=Q_7,-11Q_7=Q_3,-11Q_5=Q_5$.  Take $P=Q_3\cup Q_5\cup Q_7$. Then $-11P=P$. According to Theorem 3.7(2), $C_P$ is a  Galois LCD MDS code with parameters $[10,7,4]$.
\end{exam}

\begin{thm} Let $C_P$ is  an  $\lambda$-constacyclic code of length $n$ over $\mathbb{F}_q$  with the defining set $P$, where $\lambda^{1+p^{e-k}}=1$. For any $P\subset  (1+r\mathbb{Z}_{rn})$, $C_P$ is a Galois LCD $\lambda$-constacyclic  code if and only if there exists  some integer $j$ such that $p^{ej-k}\equiv-1~\mathrm{mod}~rn$.
\end{thm}
\pf Since $-p^{k}\equiv q^j~\mathrm{mod}~rn$ for some integer $j$, we have  $-p^{k}s\equiv q^js~\mathrm{mod}~rn$ for any $s\in P$. Thus, $-p^{k}P\subset P$.

On other hand, let $P=\cup_{i=1}^{t}Q_{s_{i}}$ for some positive integer $t$, where $Q_{s_{i}}=\{ s_i, qs_i,\ldots,q^{m_{i}-1}s_i \} \subset (1+r\mathbb{Z}_{rn})$ and $m_{i}$ is the smallest integer satisfying $q^{m_{i}}s_i \equiv s_i~\mathrm{mod}~rn $ for $i=1,\ldots,t$. For every $s_i$, we have  $-s_ip^{k}\equiv s_i q^j~\mathrm{mod}~rn$ since $-p^{k}\equiv q^j~\mathrm{mod}~rn$. Thus, $-s_ip^{k}\in Q_{s_i}$ for $i=1,\ldots,t$. Furthermore,
$s_i\equiv s_i q^{m^{i}}=s_i q^j q^{m_{i}-j}\equiv -s_ip^{k}q^{m_{i}-j}=-p^{k}(s_iq^{m_{i}-j})~\mathrm{mod}~rn$, which implies $s_i\in -p^{k}Q_{s_i}\subset -p^{k}P$.

Therefore, $-p^{k}P=P$, i.e., $C_P$ is a Galois LCD $\lambda$-constacyclic  code.

Conversely, take $P_1=\{1,q,q^2,\ldots, q^{m_{1}-1}\}$, where $m_{1}$ is the smallest integer satisfying $q^{m_1}\equiv1~\mathrm{mod}~rn$. By the assumption, $C_{P_{1}}$ is a Galois LCD $\lambda$-constacyclic  code. According to Theorem 3.7(2), we have  $-p^{k}P_{1}=P_{1}$. This means that there exists some integer $j$ such that $-p^{k}=q^{j}~\mathrm{mod}~rn$, i.e., $p^{ej-k}\equiv-1~\mathrm{mod}~rn$.
\qed
\begin{rem} It follows from Theorem 3.8 that if $rn$ divides $ 1+p^{ej-k}$,  then every $\lambda$-constacyclic code of length $n$ over $\mathbb{F}_q$ is a Galois LCD code.
\end{rem}

In light of the proof of above theorem, the following two corollaries are straightforward.
\begin{cor} If $Q_s \subset (1+r\mathbb{Z}_{rn})$  is an $q$-cyclotomic coset modulo $nr$, then $-p^{k}Q_s=Q_s$ if and only if $s(1+p^{ej-k})\equiv0~\mathrm{mod}~rn$ for some integer $j$.
\end{cor}

\begin{cor} Let $P= Q_{s}\cup (-p^{k}Q_s)$ and $-p^{k}Q_{s}\neq Q_{s}$, where $Q_{s}\subset (1+r\mathbb{Z}_{rn})$.  Then  $-p^{k}P=P$ if and only if $p^{2k}s\equiv q^js~\mathrm{mod}~rn$ for some integer $j$.
\end{cor}
\begin{thm} Let $C_P$ is  an  $\lambda$-constacyclic code of length $n$ over $\mathbb{F}_q$  with the defining set $P$, where $\lambda^{1+p^{e-k}}=1$. For any $P\subset  (1+r\mathbb{Z}_{rn})$, $C_P$ is a Galois LCD $\lambda$-constacyclic  code if and only if $Q_1=Q_{-p^k}$, where $Q_{1}=\{1,q,q^2,\ldots, q^{m_{1}-1}\}$, where $m_{1}$ is the smallest integer satisfying $q^{m_1}\equiv1~\mathrm{mod}~rn$.
\end{thm}
\pf If $Q_1=Q_{-p^k}$, then for any $s\in (1+r\mathbb{Z}_{rn}), Q_s=Q_{-p^ks}$ . For any $P\subset  (1+r\mathbb{Z}_{rn})$, we know that $P=\cup_{i=1}^{t} Q_{s_{i}}$ with $Q_{s_{i}}\in(1+\mathbb{Z}_{rn})/\mu_{q}$. Thus $-p^{k}P=P$, i.e., $C_P$ is a Galois LCD $\lambda$-constacyclic  code.

Conversely suppose for any $P\subset  (1+r\mathbb{Z}_{rn})$, $C_P$ is a Galois LCD $\lambda$-constacyclic  code. In particular, setting $P=Q_1$,  $C_{Q_1}$ is a Galois LCD $\lambda$-constacyclic  code. Therefore, $-p^kQ_1=Q_1$, which implies that $Q_1=Q_{-p^k}$.
\qed

\begin{lem} Let $p$ be an odd prime  and $n$ a positive integer such that $\mathrm{ord}_{rn}(p^{e-k})=2.$ If the group $\mathbb{Z}_{rn}^{*}$ has a unique element of order $2$, i.e.,$[-1]_{rn}$ is a unique element of order $2$ in $\mathbb{Z}_{rn}^{*}$, where  $[-1]_{rn}$ denotes the reside class modulo $rn$ containing $-1$, then, for any $s\in (1+\mathbb{Z}_{rn})$, we have $ Q_s=-p^{k}Q_s$.
\end{lem}
\pf According to the assumption that $\mathrm{ord}_{rn}(p^{e-k})=2$, from the assumption that $[-1]_{rn}$ is a unique element of order $2$ in $\mathbb{Z}_{rn}^{*}$, it follows that $p^{e-k}\equiv -1~\mathrm{mod}~rn$, i.e.,     $p^e\equiv-p^k~\mathrm{mod}~rn$. Therefore, for any $s\in (1+\mathbb{Z}_{rn})$, we have $-p^kQ_{s}=Q_{s}$.
\qed

\begin{exam} Let $p=5,e=3,i.e., q=5^3=125$. Take $k=1$, $n=13$ and $\lambda=-1$. Then $r=2,rn=26$.  Consider
 $$1+2\mathbb{Z}_{26}=\{1,3,5,7,9,11,13,15,17,19,21,23,25\}.$$
 The $q$-cyclotomic cosets modulo $26$ containing in $1+2\mathbb{Z}_{26}$ are
 $$Q_1=\{1,5,21,25\},Q_3=\{3,11,15,23\},Q_7=\{7,9,17,19\},Q_{13}=\{13\}.$$
 Since $5^{3-1}\equiv-1~\mathrm{mod}~26$,  according to Theorem 3.8 produce $15$ Galois LCD codes,
 but only  $5$ types of parameters $[13,12,2],[13,9,4],[13,8,4],[13,4,8],[13,5,7]$, respectively.
\end{exam}

\begin{exam} Let $p=13,e=3,i.e., q=13^3=2197$. Taking $k=2$, $n=9$ and $\lambda=-1$. Then $r=2,rn=18$.  Consider
 $$1+2\mathbb{Z}_{18}=\{1,3,5,7,9,11,13,15,17\}.$$
 The $q$-cyclotomic cosets modulo $18$ containing in $1+2\mathbb{Z}_{18}$ are
 $Q_1=\{1\},Q_3=\{3\},Q_5=\{5\},Q_7=\{7\},Q_9=\{9\},Q_{11}=\{11\},Q_{13}=\{13\},Q_{15}=\{15\},Q_{17}=\{17\}$.
It is easy to check that $-13^2Q_1=Q_{11},-13^2Q_{11}=Q_{13},-13^2Q_{13}=Q_{17},-13^2Q_{17}=Q_7,-13^2Q_7=Q_5,-13^2Q_5=Q_1,-13^2Q_3=Q_{15},-13^2Q_{15}=Q_3,-13^2Q_9=Q_9$. Taking $P_1=\{1,5,7,11,13,17\}$ $P_2=\{1,5,7,9,11,13,17\},P_3=\{1,3,5,7,11,13,15,17\},P_4=\{3,15\}, P_5=\{3,9,15\}$, and $P_6=\{9\}$. Then $-13^2P_i=P_i$ for $1\leq i \leq 6$. According to Theorem 3.7(2), $C_{P_{1}}$ is a Galois LCD  code with parameters $[9,3,3]$, $C_{P_{2}}$ is a Galois LCD  code with parameters $[9,2,6]$, $C_{P_{3}}$ is a Galois LCD MDS code with parameters $[9,1,9]$, $C_{P_{4}}$ is a Galois LCD  code with parameters $[9,7,2]$, $C_{P_{5}}$ is a Galois LCD  code with parameters $[9,6,2]$, $C_{P_{6}}$ is a Galois LCD  MDS code with parameters $[9,8,2]$.
\end{exam}

\section{Hermitian LCD constacyclic codes over $\mathbb{F}_{q}$}
In this section we study  constacyclic  Hermitian LCD codes over $\mathbb{F}_{q}$, where $q=p^e$ and $e=2a$. By Theorem 3.7, 3.8, 3.12,   we have the following corollary.

\begin{cor} Let $C_P$ is  an  $\lambda$-constacyclic code of length $n$ over $\mathbb{F}_q$  with the defining set $P$, where $\lambda^{1+p^{a}}=1$. For any $P\subset  (1+r\mathbb{Z}_{rn})$, $C_P$ is a Hermitian LCD $\lambda$-constacyclic  code if and only if one of the following statements holds

$\mathrm{(1)}$ $-p^{a}P=P$.

$\mathrm{(2)}$ $Q_1=Q_{-p^a}$, where $Q_{1}=\{1,q,q^2,\ldots, q^{m_{1}-1}\}$, where $m_{1}$ is the smallest integer satisfying $q^{m_1}\equiv1~\mathrm{mod}~rn$.

$\mathrm{(3)}$ There exists  some integer $j$ such that $(p^{a})^{2j-1}\equiv-1~\mathrm{mod}~rn$.
\end{cor}

By Corollary 3.3, if $\lambda^{1+p^a}\neq1$, then any $\lambda$-constacyclic  code of length $n$ over $\mathbb{F}_{p^{2a}}$ is Hermitian LCD. Then we only need to look at the  classes of  $\lambda$-constacyclic codes where $\lambda^{1+p^{a}}=1$. In this case, we gives a necessary  condition for any $\lambda$-constacyclic  code of length $n$ over $\mathbb{F}_{p^{2a}}$ to be Hermitian LCD.

\begin{thm}  Let $\lambda$ be a primite $r$th root of unity over $\mathbb{F}_{p^{2a}}$ and $r=2^{b_1}r'(b_1>0,~r'~\mathrm{odd})$. Let $n=2^{b_2}n'(b_2>0,~n'~\mathrm{odd})$, and let $q$ be an odd prime power such that $(n,q)=1$. If for any $\lambda$-constacyclic  code of length $n$ over $\mathbb{F}_{p^{2a}}$ is Hermitian LCD, then $r\mid p^a+1$ and $p^a+1\equiv0~(\mathrm{mod}~2^{b_1+b_2})$.
\end{thm}
\pf By Corollary 4.1 (3), there exists  some integer $j$ such that $(p^{a})^{2j-1}\equiv-1~\mathrm{mod}~rn$. Therefore,
$$rn\mid(p^{a})^{2j-1}+1\Rightarrow2^{b_1+b_2}r'n'\mid\frac{(p^{a})^{2j-1}+1}{p^a+1}(p^a+1).$$
Obviously, $\frac{(p^{a})^{2j-1}+1}{p^a+1}$ is odd, we have $p^a+1\equiv0~(\mathrm{mod}~2^{b_1+b_2})$.
\qed

We can check the following lemma.
\begin{lem} For some positive integer $t$, let $A=\cup_{i=1}^{t}Q_{s_{i}}$, where $Q_{s_{i}}=\{ s_i, p^es_i,\ldots,(p^e)^{m_{i}-1}s_i \} \subset (1+r\mathbb{Z}_{rn})$ and $m_{i}$ is the smallest integer satisfying $(p^a)^{m_{i}}s_i \equiv s_i~\mathrm{mod}~rn $ for $i=1,\ldots,t$. Assume that
$-p^as_i\equiv v_{1i}~\mathrm{mod}~rn,-p^as_i(p^e)\equiv v_{2i}~\mathrm{mod}~rn,\ldots,-p^as_i(p^e)^{m_{i}-1}\equiv v_{m_{i},i}~\mathrm{mod}~rn,~\mathrm{for}~i=1,\ldots,t$,
and
$P=(\cup_{i=1}^{t}Q_{s_{i}})\cup(\cup_{i=1}^{t}\{v_{1i},\ldots, v_{m_{i},i}\})$.
Then $-p^aP=P$.
\end{lem}

\begin{thm}  Let $Q_{s_{1}}, \ldots, Q_{s_{t}},Q_{w_{1}}, \ldots, Q_{w_{l}}$ be all the $p^{2a}$-cyclotomic cosets containing  in $(1+r\mathbb{Z}_{rn})$ which satisfy $-p^aQ{s_{i}}=Q{s_{i}}, i=1,\ldots,t$,  and $-p^aQ_{w_{j}}\neq Q_{w_{j}}, j=1,\ldots,l$. Then the total number of Hermitian LCD $\lambda$-constacyclic  codes of length $n$ over $\mathbb{F}_{p^{e}}$ is equal to $2^{t+h}-1$, where $l=2h$ and $h$ is a positive integer.
\end{thm}
\pf Let $-p^aQ_{w_{j}}=Q_{w_{\overline{j}}}$ and $\tau$ be a permutation of $\{1,2,\ldots,l\}$ which satisfies $\tau(j)=\overline{j}$ for $j=1,2,\ldots,l$. Then we have $Q_{w_{\tau(j)}}=Q_{w_{\tau^2(j)}}=Q_{w_{j}}$ since $p^{2a}Q_{w_{j}}=Q_{w_{j}}$. Hence, $\tau^2(j)=j$ for $j=1,2,\ldots,l$, i.e., $\tau^2=I$, where $I$ denote an identity permutation of $\{1,2,\ldots,l\}$

By assumption, we know that $\tau(j)\neq j$ for $j=1,2,\ldots,l$. Since $\tau^2=I$, $\tau$ must be a product of mutually disjoint transpositive like $(x_1y_1)\ldots(x_hy_h)$. Therefore, $l=2h$ for some integer $h$. Without loss of generality, for $j=1,2,\ldots,h$, we assume that $\tau(j)=h+j$, then  $\tau(h+j)=j$. Hence, by Lemma 4.3, the total number of Hermitian LCD $\lambda$-constacyclic  codes of length $n$ over $\mathbb{F}_{p^{e}}$ is equal to $2^{t+h}-1$.
\qed

\begin{rem} It follows from the proof of Theorem 4.4 that if $-p^aQ_{w_{j}}= Q_{w_{\overline{j}}}$ then$-p^aQ_{w_{\overline{j}}}= Q_{w_{j}}$.
\end{rem}

\begin{exam} Let $p=11,e=2,i.e., q=11^2=121$. Taking  $n=10$ and $\lambda=1$. Then $r=1,rn=10$.  Thus,
 $q$-cyclotomic cosets modulo $10$ are $Q_0=\{0\},Q_1=\{1\},Q_2=\{2\},Q_3=\{3\},Q_{4}=\{4\},Q_5=\{5\},Q_{6}=\{6\},Q_7=\{7\},Q_{8}=\{8\},Q_9=\{9\}$.
It is easy to check that $-11Q_1=Q_{1},-11Q_{2}=Q_{8},-11Q_{3}=Q_{7},-11Q_{4}=Q_6,-11Q_5=Q_5$. By Theorem 4.4, the total number of Hermitian LCD cyclic  codes of length $10$ over $\mathbb{F}_{11^{2}}$ is equal to $63$. Taking $P_1=\{4,5,6\},P_2=\{3,4,5,6,7\}$, and $P_3=\{2,3,4,5,6,7,8\}$. Then $C_{P_1},C_{P_2}$ and $C_{P_3}$ are Hermitian LCD MDS codes with parameters $[10,7,4],[10,5,6]$ and $[10,3,7]$, respectively.
\end{exam}

\begin{lem} Let $p$ be an odd prime  and $n$ a positive integer such that $\mathrm{ord}_{rn}(p^a)=2(1+2j)$ for some integer $j$. If the  group $\mathbb{Z}_{rn}^{*}$ has a unique element of order $2$, i.e.,$[-1]_{rn}$ is a unique element of order $2$ in $\mathbb{Z}_{rn}^{*}$, where  $[-1]_{rn}$ denotes the residue class modulo $rn$ containing $-1$, then,  for $0\leq i \leq n-1$, $-p^aQ_{1+ri}=Q_{1+ri}$, where $Q_{1+ri}$ is an  $p^{2a}$-cyclotomic coset modulo $rn$ containing $1+ri$ in $1+r\mathbb{Z}_{rn}$, and $\mid Q_1\mid=1+2j$.
\end{lem}
\pf Since $(p^a)^{2(1+2j)}\equiv1~\mathrm{mod}~rn$, from the assumption that $[-1]_{rn}$ is a unique element of order $2$ in $\mathbb{Z}_{rn}^{*}$, it follows that $(p^a)^{1+2j}\equiv -1~\mathrm{mod}~rn$. Therefore,  $$(p^{2a})^{1+j}\equiv -p^a~\mathrm{mod}~rn.$$
This implies that $-p^a(1+ri)\equiv(1+ri)(p^{2a})^{1+j}~\mathrm{mod}~rn$, i.e., $-p^aQ_{1+ri}=Q_{1+ri}$, for $0\leq i \leq n-1$.

As $\mathrm{ord}_{rn}(p^a)=2(1+2j)$, obviously, $\mid Q_1\mid=1+2j$.
\qed

Using the aforementioned lemma, some optimal Hermitian LCD $\lambda$-constacyclic codes of length $n$ over $\mathbb{F}_{q}$ can be constructed .

\begin{thm} Let $p$ be an odd prime  and $n$ a positive integer such that $\mathrm{ord}_{rn}(p^a)=2.$ If the group $\mathbb{Z}_{rn}^{*}$ has a unique element of order $2$, i.e.,$[-1]_{rn}$ is a unique element of order $2$ in $\mathbb{Z}_{rn}^{*}$, where  $[-1]_{rn}$ denotes the residue class modulo $rn$ containing $-1$, then, for $2\leq d \leq n$, there exists a Hermitian LCD MDS $\lambda$-constacyclic  code with parameters  $[n,n+1-d,d]$.
\end{thm}
\pf By Lemma 4.6, we have $Q_1=-p^aQ_1=\{1\}$ and $-p^a\equiv1~\mathrm{mod}~rn$. Therefore, for each $i$, $1\leq i\leq n-1$,
$$-p^a(1+ri)\equiv1+ri~\mathrm{mod}~rn.~~~~~~~~~~~~~~~~~~~~~~~~~~~~~~~(4.1)$$
According to the assumption that $\mathrm{ord}_{rn}(p^a)=2$, we obtain $$p^{2a}(1+ri)\equiv1+ri~\mathrm{mod}~rn.~~~~~~~~~~~~~~~~~~~~~~~~~~~~~~~(4.2)$$

Combing Equation $(4,1)$ and $(4,2)$, for each $i$, $1\leq i \leq n-1$, we show that $-p^aQ_{1+ri}=Q_{1+ri}=\{1+ri\}$.

Let the defining set of an $\lambda$-constacyclic code $C=\langle g(x)\rangle$  of length $n$ over $\mathbb{F}_{q}$ be the set $P=\cup_{i=0}^{d-2}Q_{1+ri}$, where $2\leq d \leq n$. Then $-p^aP=P$. By Corollary 4.1, the code $C$ is a Hermitian LCD $\lambda$-constacyclic  code and $\mathrm{dim}C=n+1-d$.  Obviously, the defining set $P$ consists of $d-1$ consecutive  integers $\{1,1+r,1+2r,\ldots,1+(d-2)r\}$. Using Theorem 3.5, the minimum distance of $C$ is at least $d$. Thus, we conclude that $C$ is an $\lambda$-constacyclic Hermitian LCD code with parameters  $[n,n+1-d,\geq d]$.  Applying the classical code Singleton bound to $C$ yields a  Hermitian LCD MDS $\lambda$-constacyclic code with parameters  $[n,n+1-d,d]$.
\qed

\begin{exam} Let $p=3, a=2,r=2$, and $n=5$. Then $p^{2a}\equiv 1~\mathrm{mod}~10$ and $\mathbb{Z}_{10}^{*}=\{1,3,7,9\}$ has a unique element $9$ of order $2$.  Applying Theorem 4.7 produce $9$ Hermitian LCD MDS negacyclic  codes with parameters $[5,1,5],[5,2,4],[5,3,3],[5,4,2]$,respectively.
\end{exam}

\section*{Acknowledgements}
The research of the authors is supported by NSFC
with grant numbers 11271005. Hualu Liu was supported by China Scholarship Council (Grant No. 201606770024), the excellent
doctorial dissertation cultivation grant from Central China Normal
University(Grant No. 2016YBZZ082), and the Educational Commission of Hubei Province (Grant No. B2015096).

\end{document}